\documentclass[12pt]{article}
\usepackage{times}
\usepackage{epsfig,amsmath}
\usepackage{subfigure}
\usepackage{graphicx}
\usepackage{color}
\usepackage{cite}

\usepackage{epstopdf}

\newenvironment{sciabstract}{%
\begin{quote} \bf}
{\end{quote}}

\newcounter{lastnote}

\title{Two-Dimensional Quantum Walk of Correlated Photons}

\author
{Zhi-Qiang Jiao,$^{1,2\dagger}$ Jun Gao,$^{1,2\dagger}$ Wen-Hao Zhou,$^{1,2}$ Xiao-Wei Wang,$^{1,2}$\\
Ruo-Jing Ren,$^{1,2}$ Xiao-Yun Xu,$^{1,2}$ Lu-Feng Qiao,$^{1,2}$ and Xian-Min Jin$^{1,2\ast}$\\
\\
\normalsize{$^1$Center for Integrated Quantum Information Technologies (IQIT), School of Physics}\\
\normalsize{and Astronomy, State Key Laboratory of Advanced Optical Communication Systems}\\
\normalsize{ and Networks, Shanghai Jiao Tong University, Shanghai 200240, China}\\
\normalsize{$^2$CAS Center for Excellence and Synergetic Innovation Center in Quantum Information and}\\
\normalsize{Quantum Physics, University of Science and Technology of China, Hefei, Anhui 230026, China}\\
\normalsize{$^\dagger$These authors contributed equally to this work}\\
\normalsize{$^\ast$E-mail: xianmin.jin@sjtu.edu.cn}\\
}

\date{}
\begin{document}
\baselineskip24pt

\maketitle

\begin{sciabstract}

Quantum walks in an elaborately designed graph, is a powerful tool simulating physical and topological phenomena, constructing analog quantum algorithms and realizing universal quantum computing. Integrated photonics technology has emerged as a versatile platform to implement various quantum information tasks and a promising candidate to perform large-scale quantum walks. Both extending physical dimensions and involving more particles will increase the complexity of the evolving systems and the desired quantum resources. Pioneer works have demonstrated single particle walking on two-dimensional (2D) lattices and multiple walkers interfering on a one-dimensional structure. However, 2D multi-particle quantum walk, genuinely being not classically simulatable, has been a vacancy for nearly ten years. Here, we present a genuine 2D quantum walk with correlated photons on a triangular photonic lattice, which can be mapped to a state space up to 37$\times$37 dimensions. This breaks through the physically restriction of single-particle evolution, which can encode information in a large space and constitute high-dimensional graphs indeed beneficial to quantum information processing. A site-by-site addressing between the chip facet and the 2D fanout interface enables an observation of over 600 non-classical interferences simultaneously, violating a classical limit up to 57 standard deviations. Our platform offers a promising prospect for multi-photon quantum walks in a large-scale 2D arrangement, paving the way for practical quantum simulation and quantum computation beyond classical regime.

\end{sciabstract}

\indent 
Random walks is a versatile mathematic tool utilized in a broad range from physics, economics to computer science. Quantum walks (QWs)\cite{Aharonov1993, Farhi1998}, the quantum extension of classical random walks, even provides an exponential speed-ups over classical random walks for certain problems due to its inherent quantum superposition\cite{Childs2003}. This unique feature leads QWs an advanced approach to building quantum algorithms\cite{Childs2003, Childs2004, algorithm1, algorithm2, Franco2011}, simulating various transport processes, like energy transport in photosynthesis\cite{Energy_transfer}, light-harvesting molecule\cite{biomolecules}, perfect state transfer\cite{state_transfer}, coherent transport\cite{coherent_transport, coherent_transport2}, and realizing universal quantum computation\cite{universal_quantum_computing1, universal_quantum_computing2}. A profusion of QWs experiments are implemented in diverse platforms, such as nuclear magnetic resonance\cite{Du}, trapped atoms and ions\cite{atom, ion}, superconducting systems\cite{superconduct}, fibers\cite{fiber, Defienne2016} and especially integrated photonic systems\cite{QW_waveguide, circular_waveguide, 2photon_QW, 2photon_cross, Bosonic-Fermionic, 2DQW, fast_hitting} for the robust and coherent nature of photons. 

An additional dimension can be conducive to simulating unexplored physical phenomena in lower dimensions and solve computationally hard problems. As shown in Fig.\ref{fig1}A, efforts have been made to introduce physical and synthetic dimensions including spatial mode, temporal loop, and momentum space\cite{quadratic_waveguide, QW_Silberhorn, Longhi, Bloch_oscillation, Pan2018, Errico2020, 4D_Hall, topological_insulator, synthetic_dimension, gauge}, which still suffer certain limitations if merely relied on the increase of dimensions. For example, spatial dimensions are hard to expand beyond 3D, and intrinsic light leakage hampers the temporal modes to beat a sufficiently large scale. Despite the ballistic and transient natures demonstrated in these systems, dynamics of single particle is simulatable by classical light, showing no genuine quantum features\cite{classical_theory}. An alternative way is to involve more walkers. The quantum interference or interaction among individual quantum walkers will introduce non-classical correlations, which is distinctly different from single walker\cite{2photon_QW, 2photon_cross, Bosonic-Fermionic} and cannot be simulated by classical light\cite{classical_theory, 2DQW}. In general, these quantum features exponentially expand Hilbert space with a linear increase of quantum walker number\cite{2photon_QW, 2photon_cross}, and can exhibit effective speed-ups and quantum advantages over coherent light. 

Simultaneously enlarging photon numbers and extending physical dimensions can provide a scalable route to increasing graph complexity and connectivity. However, the experimental implementation of such scheme remains challenging due to the absence of single-site-addressing detection for tens of sites in 2D geometry. There have been constant endeavors to push forward the developments of 2D QWs with multiple photons. Previous attempt of QWs in a physical quasi 2D, `Swiss cross' arrangement\cite{2photon_cross}, is achieved by a photonic lattice inscribed by femtosecond laser direct-writing technique. Such structure only takes one column in vertical orientation into consideration and doesn't fully utilize the 2D structure, limiting walkers moving freely along all directions. Femtosecond laser direct-writing technique is particularly suitable for engineering 2D arrangements\cite{fs-laser_writing}, and recently a large 2D lattice, up to 49$\times$49 modes, was realized by this technique to implement single-photon QW\cite{2DQW}.

Here, we experimentally demonstrate 2D QWs using correlated photons mapping to a graph with extraordinarily high dimension and connectivity. The cross-section of triangular lattice is precisely prototyped to couple to a 2D fanout interface. By injecting two indistinguishable photons into the 2D lattice, we observe distinct quantum interference and quantum correlations that strongly violate Cauchy-Schwarz inequality\cite{certification}. Our work provides a paradigm to construct a general large-scale and high-dimensional optical platform suitable for practical quantum simulation and quantum computing.  

The 2D photonic lattice is fabricated using 3D femtosecond laser direct writing technique (see Methods for fabrication details). The separation between two adjacent sites in the coupling zone is 15$\mu m$. After the interaction zone, the separation is adiabatically expanded to 35$\mu m$ while maintaining the same structure geometry to match the 2D fanout layout, as Fig.\ref{fig2}C inset shows. 35$\mu m$ is sufficient large to isolate photon hopping between adjacent sites. The cross-section of the lattice is depicted in Fig.\ref{fig1}B, which is also the state space of single photon populating $N=37$ lattices. Each site in the triangular lattice has six edges and homogenous coupling to the neighboring sites, providing more degrees of connectivity compared to the square lattices. We illustrate the connected graph corresponding to two-photon injection scenarios for 2 to 4-layer triangular lattice in Fig.\ref{fig1}D, Fig.\ref{fig1}E and Fig.\ref{fig1}F, respectively. The enlarged state space increases exponentially with the number of injected particles. The largest connected graph contains 1369 sites and 6600 edges. The edges display the hopping and transition between different basis states. All sites in the graph represent the state space of two-photon populations, and red ones are particularly marked to highlight the bunching effect, as the enlarged area in Fig.\ref{fig1}C shows.

Dynamics of single photon propagating through $N=37$ triangular lattice can be described by Hamiltonian\cite{Hamiltonian} (see Methods), 
\begin{equation}
H = \sum_{i=1}^{N}\beta_{i}a_{i}^{\dagger}a_{i} + \sum_{i\neq j=1}^{N}C_{i,j}a_{i}^{\dagger}a_{j}.
\label{H}
\end{equation}
$H$ is equivalent to the adjacency matrix of the connected graph (see Fig.\ref{fig1}B), where $\beta_{i}$ is the propagation constant of site $i$, the coupling strength between different site $i$ and $j$ is $C_{i,j} = C_{j,i}$, $a_{i}^{\dagger} ~(a_{i})$ is the bosonic creation (annihilation) operator for site $i$. In experiment, all propagation constants are engineered the same $\beta_{i} = \beta$ and coupling strength between adjacent sites is identical $C_{i,j} = C$. The dynamic evolution of single photon over length $z$ is determined by the unitary transform $U=exp(-iHz)$. The single-photon transition probability from site $i$ to $j$ is given by $p_{i, j}(z)=\left|U_{i, j}(z)\right|^{2}$, which can also be reproduced by the photon density distribution with coherent light.

The experimental setup is schematically shown in Fig.\ref{fig2}A, B and C, which represent quantum state preparation, on-chip unitary operation and large-scale coincidence measurement. We use a 780$nm$ Ti: Sapphire solid state femtosecond laser pumping a $LiB_{3}O_{5}$ (LBO) crystal to generates up-converted 390$nm$ ultraviolet pulses. Then the 390$nm$ pulses pump a $\beta-BaB_{2}O_{4}$ (BBO) crystal fulfilling type-II phase-matching to generate correlated photons in a Beam-like scheme\cite{Beamlike}. The correlated photons are filtered by 3$nm$ bandpass filter to ensure spectrum indistinguishability. Temporal overlap is achieved by an external delay line with a motorized translation stage in one arm of setup. After proper polarization compensation, correlated photons are coupled into the photonic chip by a 20X objective in free space. Through accurate and precise alignment, the coupling efficiency for both input ports can reach a balance over 50$\%$. 

We first inject coherent light into port -1 and 1 respectively, and the output pattern of the chip is accumulated by a charge-coupled device (CCD). The photon probability distribution can be extracted from the accumulated pattern, which has been extensively adopted in previous coherent light and single-photon experiments. The experimental results are illustrated in Fig.\ref{fig2}F and G compared with theoretical simulations displayed in Fig.\ref{fig2}D and E. We calculate the similarity between the patterns as $S_{i} = (\sum_{j}\sqrt{p_{i,j}^{exp}\cdot p_{i,j}^{th}})^{2}/(\sum_{j}p_{i,j}^{exp}\sum_{j}p_{i,j}^{th})$ , and the results are up to 99.6$\%$ and 98.4$\%$ respectively. 

We reconstruct the probability distributions with single-photon injection instead of coherent light, as shown in Fig.\ref{fig3}A and B. The distributions show no great differences from the results in Fig.\ref{fig2}. As for the two-photon injection scenario, the probability distribution cannot reveal any quantum feature since the result is merely an incoherent sum. To reveal quantum correlations that occur due to multi-particle interferences, two-photon correlation function $\Gamma_{i^{\prime}, j^{\prime}}^{(i, j)}$ is introduced\cite{Mattle1995},
\begin{equation}
\Gamma_{i^{\prime}, j^{\prime}}^{(i, j)}=\frac{1}{1+\delta_{i^{\prime}, j^{\prime}}}\left|U_{i^{\prime}, i}(z) U_{j^{\prime}, j}(z)+U_{i^{\prime}, j}(z) U_{j^{\prime},i}(z)\right|^{2}.
\label{cof}
\end{equation}
where $i,j$ are the input ports while $i^{\prime}, j^{\prime}$ are the output ports. The correlation function reveals genuine quantum features and bosonic bunching, see Fig.\ref{fig3}C. Therefore, it is experimentally necessary to conduct coincidence measurements and enumerate all combinations between different sites. For 1D waveguide array, this issue can be solved by coupling the photonic chip to a V-grooved fiber array with designed spacing, for example 127$\mu m$. However, the situation becomes extremely challenging for 2D QWs. Previous works transform the 2D structure into 1D alignment to match the fiber array\cite{2photon_cross}, but inevitably introduce differential losses for different channels and may cause random couplings during the transforming process. We overcome this bottleneck by directly mapping the 2D cross-section to a 2D fanout interface (see Extended Data Fig.1) and connecting the output fibers to avalanche photodiode (APD) array (see Fig.\ref{fig2}A). A homemade multi-channel coincidence module (MCCM) allows recording and processing of large-scale coincidence counting measurements. 
   
We measure the complete quantum correlation matrix in case where correlated photons are injected into the sites -1 and 1. The degree of indistinguishability is tuned by varying the relative temporal delay. The MCCM simultaneously records all $\left(\begin{array}{c}37 \\ 2\end{array}\right)=666$ HOM interference curves\cite{HOM}. We obtain the zero-delay position by fitting one of HOM interference curves. The data are collected for 2,000s at the zero-delay position to retrieve the complete quantum correlation matrix. The classical correlation can be readily calculated by classical probability theory as $\Gamma_{i^{\prime}, j^{\prime}}^{(i, j)(c)}(z)=p_{i^{\prime}, i}(z) p_{j^{\prime}, j}(z)+p_{i^{\prime}, j}(z) p_{j^{\prime}, i}(z)$. Fig.\ref{fig4}A and C show the theoretical probabilities of distinguishable and indistinguishable photons, respectively. We list the measured 2-photon coincidence correlation matrix of distinguishable and indistinguishable photons in Fig.\ref{fig4}B and D. From the experimental data, we can see that D appears obvious differences in green and yellow columns with B as the values increase distinctly, implying quantum features stemming from quantum interference. We calculate the matrix similarity $S = (\sum_{i,j}\sqrt{\Gamma_{i,j}^{exp}\cdot\Gamma_{i,j}^{th}})^{2}/(\sum_{i,j}\Gamma_{i,j}^{exp}\sum_{i,j}\Gamma_{i,j}^{th})$ to characterize the discrepancy. Finally the experiment results reveal a similarity of 91.8$\%$ with regard to the simulation. The discrepancy comes from the imperfection of indistinguishability and differential coupling efficiency. The average pitch error of the 2D fanout is about 1$\mu m$, which may cause slight coupling loss for certain channels.
     
The bunching effect is further measured by a balanced fiber beam splitter connected to the 2D fiber array to resolve the number of photons in the same site. Fig. \ref{fig5}A shows the measured peak with a visibility up to 92$\%\pm$3.5$\%$ for site 0. The Cauchy-Schwarz inequality sets a stringent bound for classical light field, and a violation of the inequality indicates a clear quantum phenomenon. Diagonal correlations $\Gamma_{i,i}$ are related to correlations in the off-diagonal $\Gamma_{i,j}$, $i \neq j$ according to the inequality\cite{certification}:
\begin{equation}
V_{i,j} = \frac{2}{3}\sqrt{\Gamma_{i,i}^{c}\Gamma_{j,j}^{c}} - \Gamma_{i,j}^{c} < 0
\label{violation}
\end{equation}
with $\Gamma^{c}$ here referring to intensity correlations between classical light beams. The nonclassical nature of the measured correlations can be quantified by the violations of Cauchy-Schwarz inequality\cite{certification}. We observe violations of classical limits spreading over the 2D lattice, as depicted in Fig.\ref{fig5}B. The largest departure of standard deviations reaches 57, manifesting strong quantum features. 

To conclude, we have experimentally demonstrated a genuinely spatial 2D QWs of correlated photons and verified multi-particle quantum effects. We observe a strong violation of Cauchy-Schwarz inequality Eq.(\ref{violation}) from quantum correlations that cannot be simulated in classical systems. Our experiment surmounts the challenge of site-by-site addressing 2D interface, which enables a direct observation of large-scale coincidence measurements. Our 3D photonic chip inscribed by direct laser writing implements a sophisticated topological geometry for manipulating multiple photons in a large 2D lattice. This genuinely scalable 2D arrangement associated with multi-photon walkers is elegantly suitable for constructing analog quantum algorithms, for example, Grove algorithms\cite{Franco2011} or exhibiting quantum supremacy\cite{supremacy} over classical supercomputers. A series of applications of QWs in higher dimensions deserve further investigations beyond the standard model, including studying the effect of multi-photon localization in higher dimensions, the dynamics and transport properties of complex topologies of multiple photons. Other quantum states, such as entanglement sates or N00N state, interacting in a topological system with multi-dimensions beyond physical limit, remains to be investigated in 2D boundary conditions. Our scalable experimental architecture is significantly conducive to the development of large-scale analog quantum computing or the near-term Noisy Intermediate-Scale Quantum(NISQ)\cite{NISQ} technologies.
 
\subsection*{Acknowledgments.}
The authors thank Jian-Wei Pan for helpful discussions. This work was supported by National Key R\&D Program of China (2019YFA0308700 and 2017YFA0303700); National Natural Science Foundation of China (NSFC) (61734005, 11761141014, 11690033); Science and Technology Commission of Shanghai Municipality (STCSM) (17JC1400403); Shanghai Municipal Education Commission (SMEC) (2017-01-07-00-02-E00049); X.-M. J. acknowledges additional support from a Shanghai talent program.

\subsection*{Data availability.}
The data that support the findings of this study are available from the corresponding author upon reasonable request.
\\

\subsection*{Methods}

\begin{figure*}[!t]
\centering
\includegraphics[width=1\linewidth]{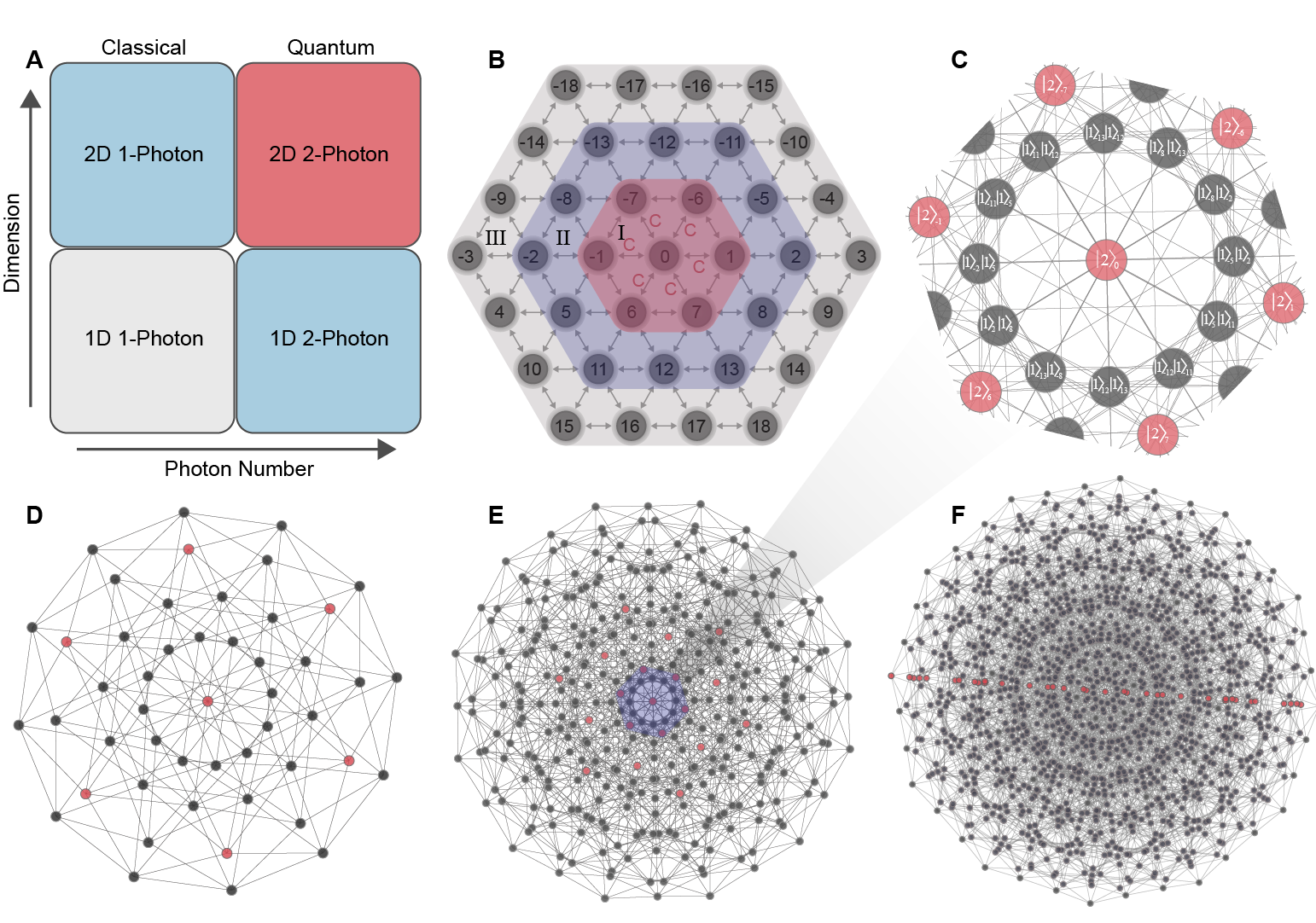}
\caption{\textbf{Architecture for QWs and schematic of high-dimensional graph structures.} \textbf{A.} Both dimensions and photon numbers can increase the complexity of photonic systems and the desired quantum resources. Single particle QWs are under description of classical wave theory. When taking more than one indistinguishable photon, QWs will step into quantum regime. One photon 1D, two photon 1D and one photon 2D have all been achieved, leaving 2D two-photon QWs (red region) unexplored. \textbf{B.} Graph structure of single-photon population. The state space of single-photon population shares the same structure with cross section of the 2D lattice. The coupling strength $C$ is uniform for all adjacent sites. The whole graph contains three layers I, II and III. \textbf{C.} Enlarged central part of high-dimensional graph\textbf{E}. The site label represents the photon basis states. The red ones are two-photon bunching states while the gray ones mean ordinary two-photon states. \textbf{D.}, \textbf{E.}, and \textbf{F.} Sketch of high-dimensional graph structures spanned by a two-photon state injection, corresponding to 7 sites (I), 19 sites (II), 37 sites(III) respectively. The graph complexity and connectivity grows exponentially with the photon number increase. Each site indicates a two-photon state with multiple edges showing flexible degrees of transitions. Every edge represents a possible transition between two-photon state, which corresponds to a single photon injecting into a higher dimensional graph structure. \textbf{F.} is the experimentally demonstrated structure, which contains 1369 sites and 6600 edges.}
\label {fig1}
\end{figure*}

\begin{figure}[!t]
\centering
\includegraphics[width=0.79\linewidth]{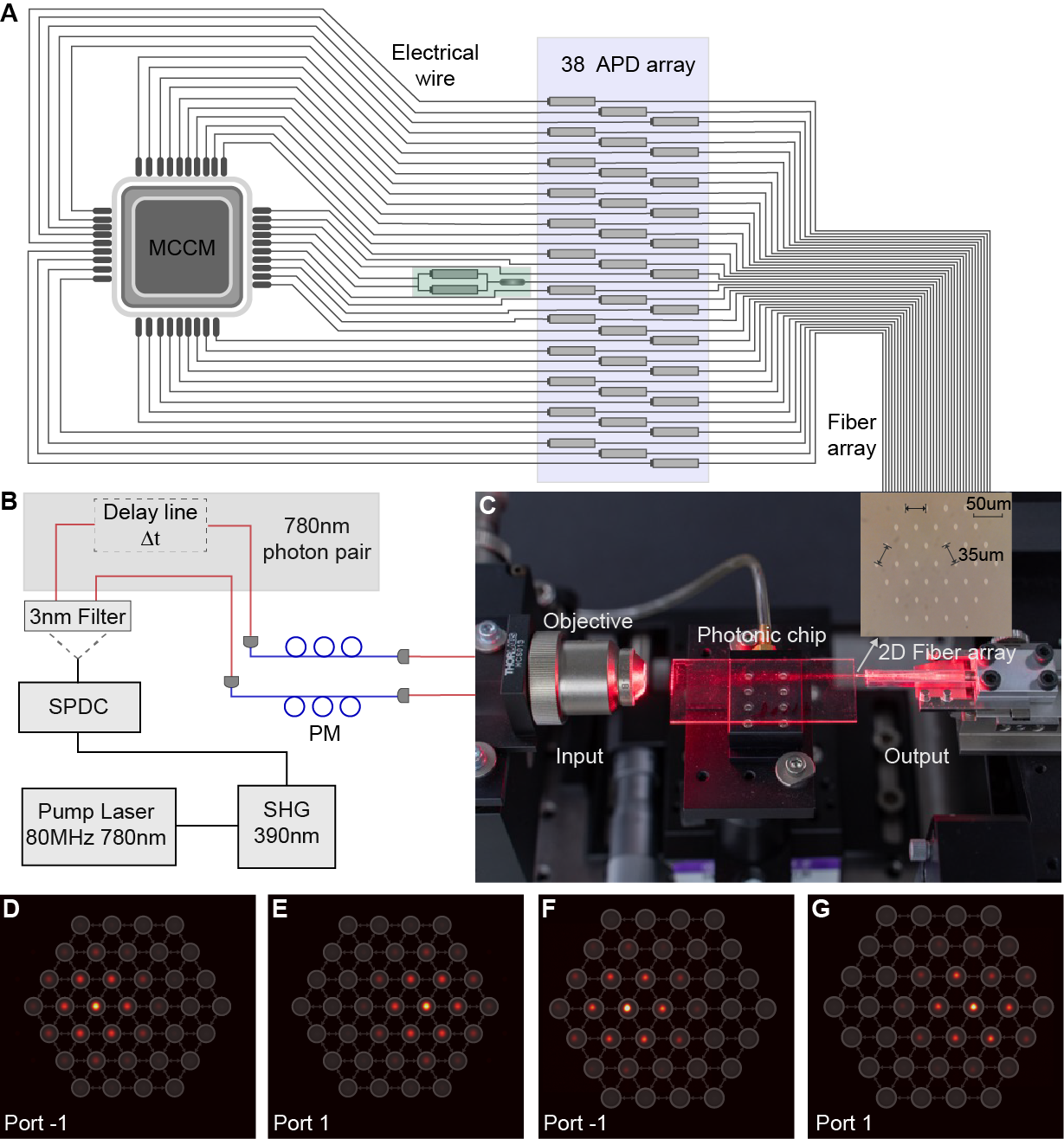}
\caption{\textbf{Sketch of experimental implementation of 2D QWs of correlated photons.} \textbf{A.} Diagram of detection system. There are 38 APDs employed in experiment to detect single-photon signals. A homemade multi-channel coincidence module (MCCM) links the detection system, simultaneously recording all the single clicks and combinations of coincidences between arbitrary two channels. Green zone shows bunching events detected by a balanced fiber beam splitter to achieve multiplxed detections. \textbf{B.} Correlated photon pair generation. A 780$nm$ femtosecond laser pumps a $LiB_{3}O_{5}$ (LBO) crystal to generates a 390$nm$ pulse by Second-harmonic generation (SHG). Then the 390$nm$ UV pulse pumps a $\beta-BaB_{2}O_{4}$ (BBO) crystal via type-II phase-matching and generates photon pairs in a Beam-like scheme. The correlated photons are filtered by 3$nm$ bandpass filters. An external delay line in one arm of the setup tunes distinguishability of photons. \textbf{C.} A 20X objective couples the correlated photons into the photonic chip in free space. The inset shows the output 2D cross-section of the chip. A 2D fanout are connected to the 2D photonic chip to achieve site-by-site couplings. \textbf{D.} (\textbf{E.}) Theoretical simulation of particle density of photons injected into port -1 (1). \textbf{F.} (\textbf{G.}) Experimental intensity distribution of coherent light injection into port -1 (1) accumulated by a CCD.}
\label {fig2}
\end{figure}

\begin{figure}[htbp]
\centering
\includegraphics[width=0.85\linewidth]{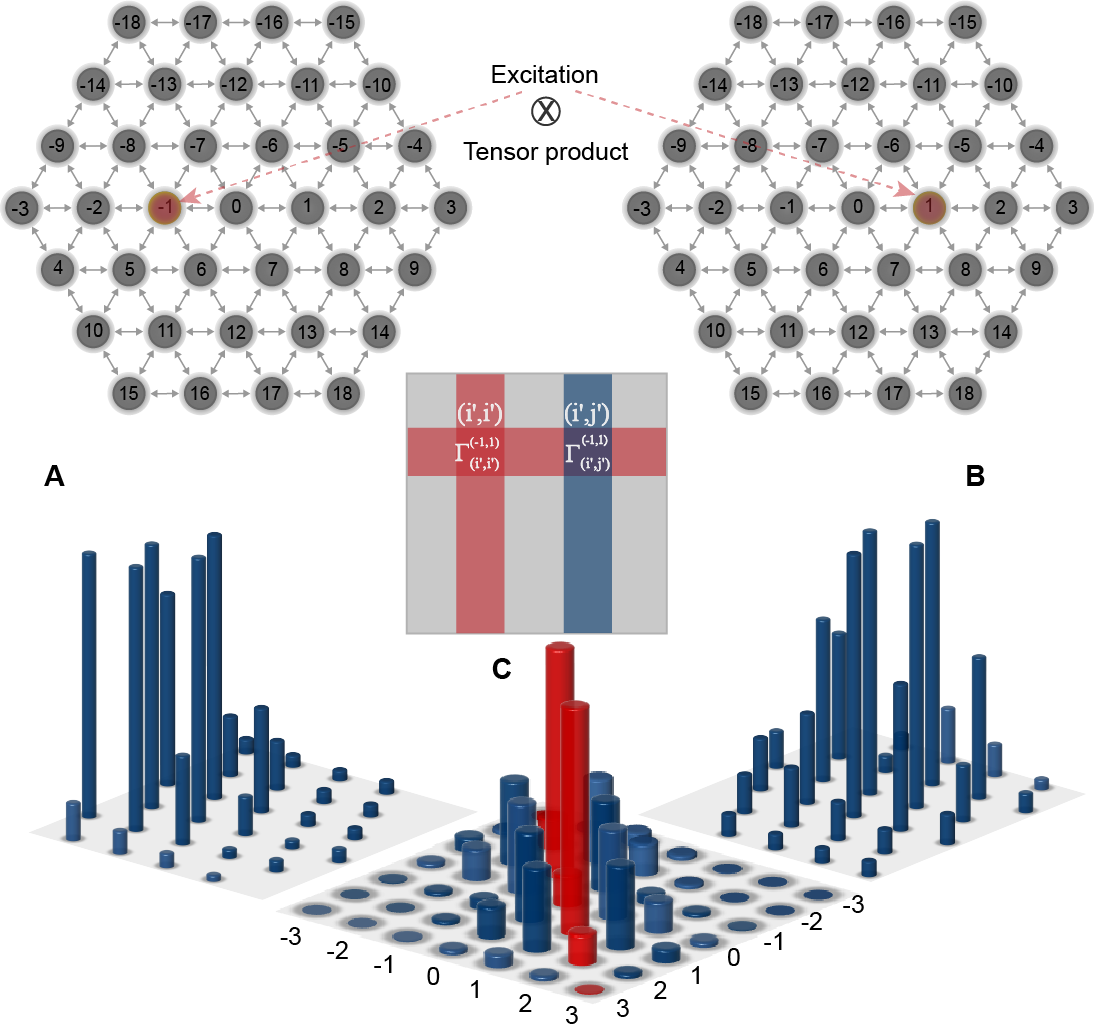}
\caption{\textbf{Illustration of measuring correlation matrix of two-photon injection.} \textbf{A.} (\textbf{B.}) Single-photon probability distributions measured by the APD detection system for port -1 (1). \textbf{C.} Two-photon correlation function $\Gamma_{i^{\prime}, j^{\prime}}^{(i, j)}$ is introduced to reveal quantum correlations. Parts of correlation matrix are listed to demonstrate 2D QWs of correlated photons. The red columns mean two-photon bunchings in the same site and the blue ones are correlations between different sites. Height indicates the degree of correlation. }
\label{fig3}
\end{figure}

\begin{figure}[htbp]
\centering
\includegraphics[width=0.9\linewidth]{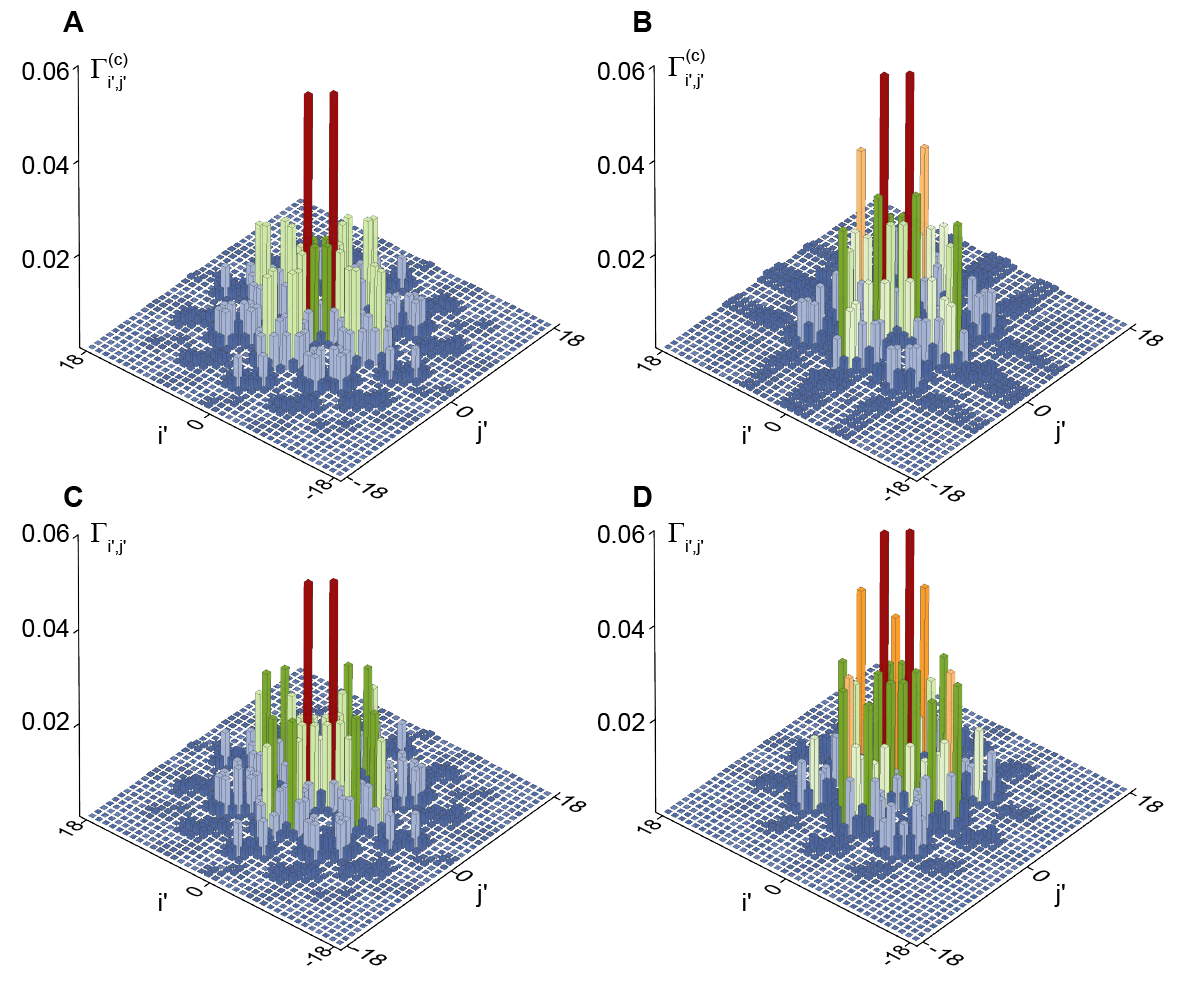}
\caption{\textbf{Simulated and measured correlations matrix.} \textbf{A.} and \textbf{B.} reveal the theoretical and experimental correlation matrix of distinguishable photons coupled to sites -1 and 1, respectively.  \textbf{C.} and \textbf{D.} illustrate the theoretical and experimental correlation matrix obtained from indistinguishable photons injected into site -1 and 1. \textbf{B.} and \textbf{D.} exhibit clear differences in green and yellow columns, implying distinct quantum features stemming from quantum interference.}
\label{fig4}
\end{figure}

\begin{figure}[htbp]
\centering
\includegraphics[width=0.98\linewidth]{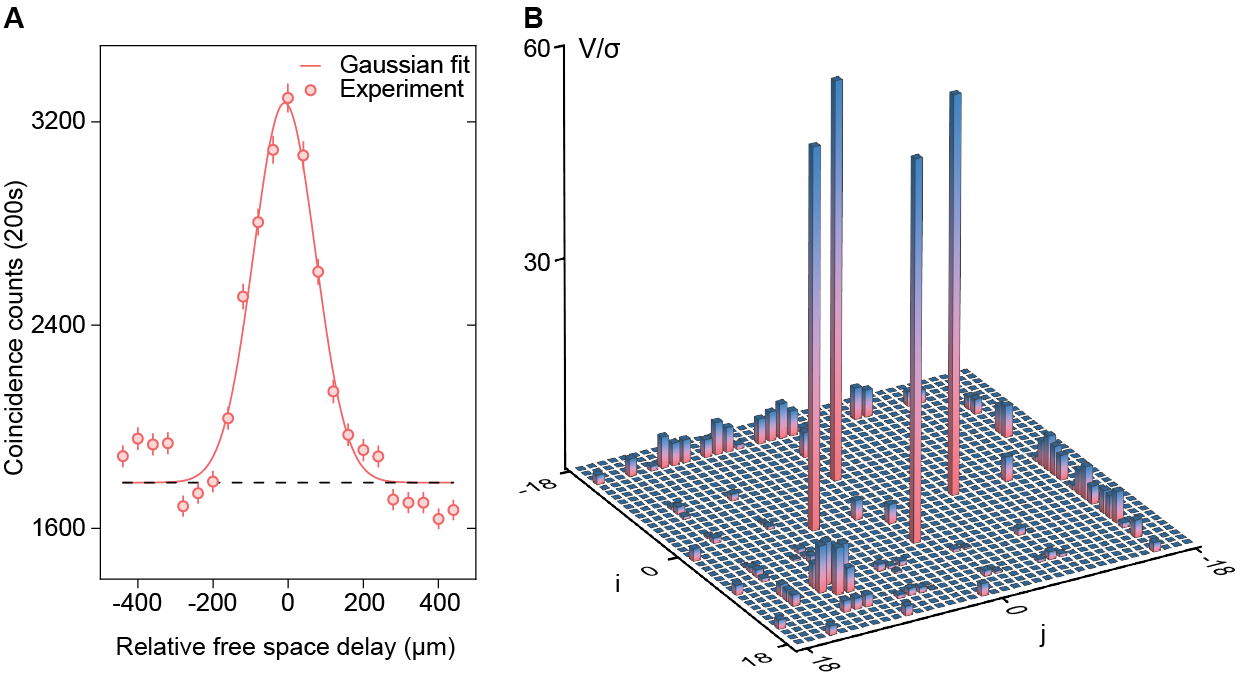}
\caption{\textbf{Non-classical certification of two-photon QWs.} \textbf{A.} Photon bunching curve retrieved from one site. By continuously shifting the relative free-space delay between these two photons, distinct quantum interference peak is observed for site 0. This typical interference curve shows a visibility of 92$\% \pm$ 3.5$\%$. \textbf{B.} Violation of Cauchy-Schwarz inequality. These violations are certified from the standard deviations $\sigma$, assuming individual counts following Poissonian statistics. The histograms with red color represent violation and vice versa. The maximum violation reaches 57 standard deviations. Only indistinguishable photons show bosonic bunching features that enable the appearance of violations as expected.} 
\label {fig5}
\end{figure}

\paragraph{2D photonic triangular lattice fabrication:}
Focusing femtosecond-laser pulses into a borosilicate glass (Eagle XG) sample permanently modifies the material in the focal volume, resulting in a refractive index increase. A second harmonic generation of the femtosecond laser system supplies 290$fs$ pulse at a central wavelength of 513$nm$ with a repetition rate of 1$MHz$. We feed the laser into a cylindrical lens to reshape the beam into a narrow one, and then focus the laser beam by a 50X objective lens (0.55NA) into the 5 $cm$ long substrate to inscribe waveguides array through moving the substrate held on a high-precision air-bearing stage. We fix the laser power to 210$nJ$ and a constant writing velocity of 15$mm/s$. The middle layer waveguides lie in the depth of 170$\mu m$ below the surface. The injection port distance is 130$\mu m$ with a bending radius of 30$mm$. The coupling zone shares the same structure as Fig.\ref{fig1}(B) and has an evolution length of 11$mm$. We adiabatically expand the waveguide pitch of 15$\mu m$ in the coupling zone to match the 2D fanout pitch of 35$\mu m$ in an adiabatic length of 4$mm$, as Extended Data Fig.1 shows. 

\paragraph{Coupling strengths characterization:}
We inject horizontal 780$nm$ laser into waveguide to characterize a series of coupling strengths shown in Extended Data Fig.2. The coupling strength displays an exponential decay along the separation between two sites. In our experiment, we choose 15$\mu m$ for the waveguide spacing. For simplicity we assume all coupling strengths in the triangular structure to be uniform and so is the propagation constant. In practice, deviations may emerge during fabrication process and will introduce asymmetries in the triangular lattice. The good similarity of single-photon evolution patterns between experiments and simulations indicates the above asymmetries are negligible.

\paragraph{Continuous quantum walk on a chip:}
In a photonic lattice, the mode fields of neighboring single-mode waveguides are overlapped and light in waveguides can experience quantum tunneling. Photons propagating through evanescently coupled waveguides is defined by the following Hamiltonian,
\begin{equation}
H = \sum_{i=1}^{N}\beta_{i}a_{i}^{\dagger}a_{i} + \sum_{i\neq j=1}^{N}C_{i,j}a_{i}^{\dagger}a_{j}.
\label{H}
\end{equation}
where $H$ is equivalent to the adjacency matrix of the connected graph, $\beta_{i}$ is the propagation constant of site $i$, and the coupling strength between different sites $i$ and $j$ is $C_{i,j} = C_{j,i}$. In general, we only consider quantum tunneling that happens between one site and its nearest neighbors. For example, port 0 has 6 possible trajectories. $a_{i}^{\dagger} ~(a_{i})$ is the bosonic creation (annihilation) operator for site $i$. The propagation dynamics of a single photon is described by Heisenberg equation of motion, $da^{\dagger}(z)/dz = -i[a^{\dagger},H]/\hbar$. We replace the evolution time $t$ by length $z=ct$, where $c$ is the light speed inside waveguide. Since the Hamiltonian is time independent, we can simplify the unitary evolution operator to be $U=exp(-iHz)$. The solution of the equation regarding the operator is $a_{j}^{\dagger}(z) = \sum_{i=1}^{N}U_{j,i}(z)a_{i}^{\dagger}(0)$ by applying unitary operator on the input mode operator $a_{i}^{\dagger}$. The single-photon evolution distribution is calculated by the average photon number $n_{j}=\left\langle a_{j}^{\dagger} a_{j}\right\rangle$ in site $j$. Coincidence measurements can be described via quantum correlation function,
\begin{equation}
\Gamma_{i, j}(z)=\left\langle\psi(0)\left|a_{i}^{\dagger}(z) a_{j}^{\dagger}(z) a_{j}(z) a_{i}(z)\right| \psi(0)\right\rangle.
\end{equation}

\clearpage
\renewcommand\refname{References and Notes}

\clearpage

\clearpage

\clearpage

\clearpage

\clearpage

\begin{figure}
\centering
\includegraphics[width=0.95\linewidth]{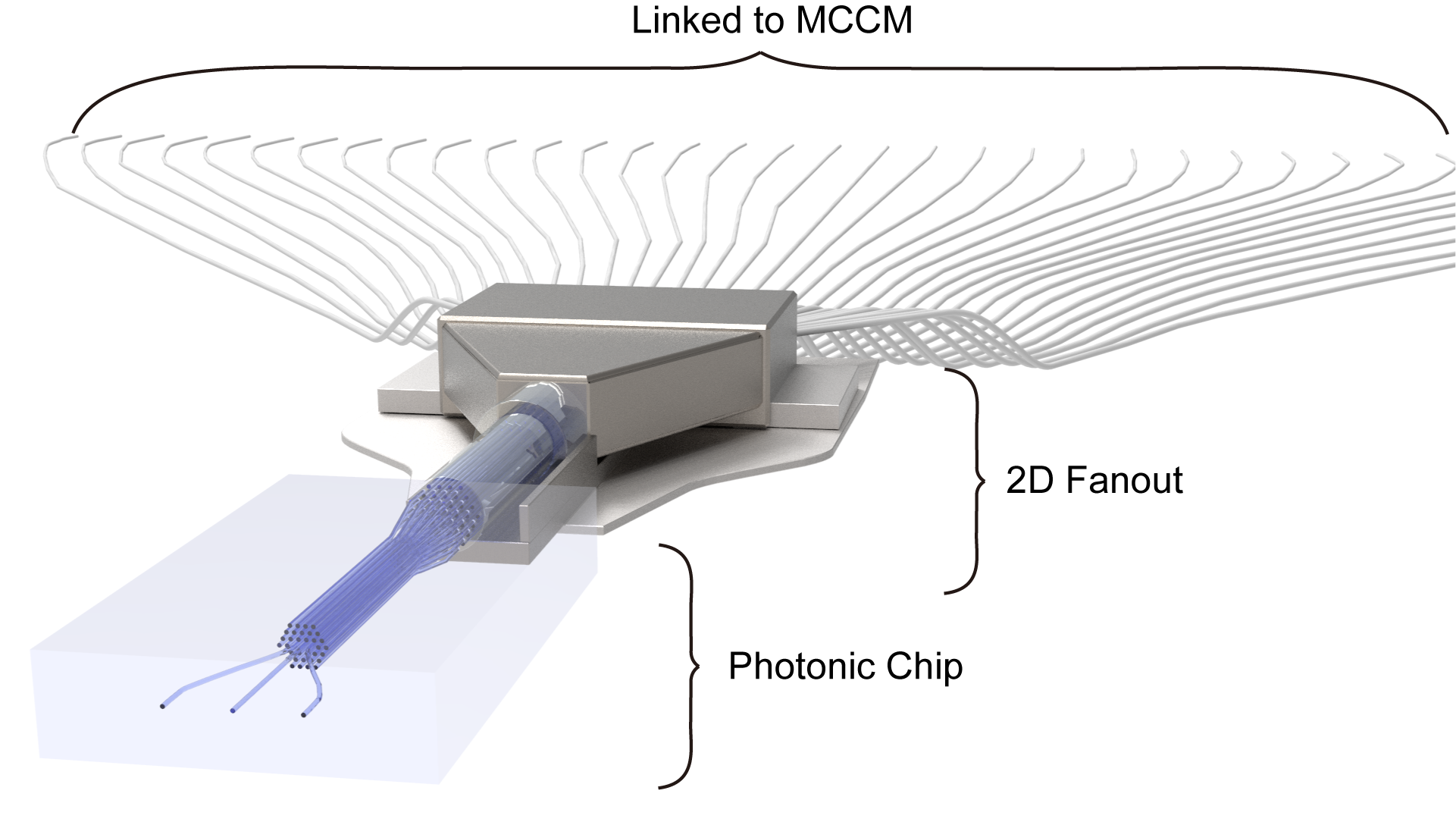}\\
\flushleft
Extended Data Fig.1 \textbf{Sketch of 2D fanout interface connected to the 2D photonic lattice.}  The photonic lattice has 3 input ports and 37 sites in the coupling zone with 15$\mu m$ spacing. The coupling zone expands adiabatically to match the 2D fanout interface. All fibers are connected to APDs and MCCM records all coincidence measurements simultaneously.
 \label{fig6}
\end{figure}

\begin{figure}
\centering
\includegraphics[width=0.55\linewidth]{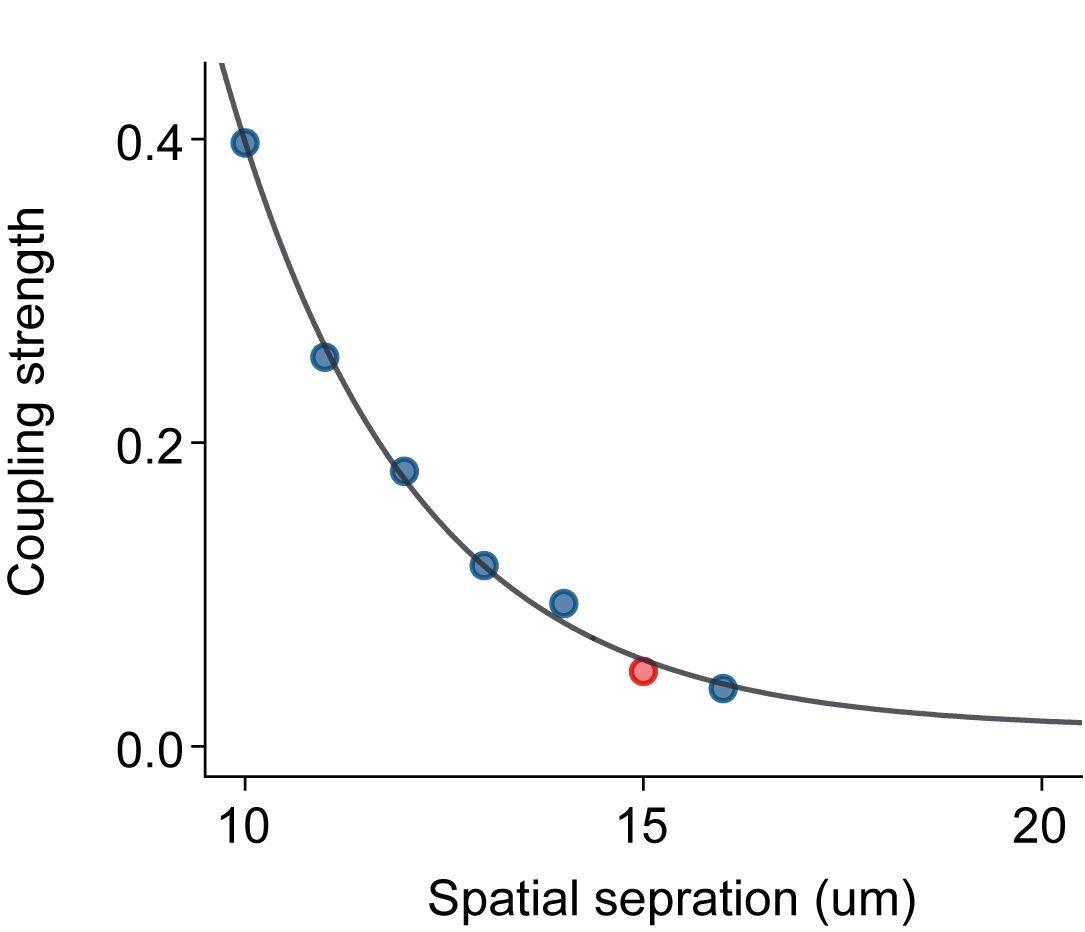}\\
\flushleft
Extended Data Fig.2 \textbf{Coupling strength characterization. } Experimental characterization of coupling strength with different spatial separations. The coupling strength curve displays an exponential decay. In our experiment, the spacing in the coupling zone is chosen as 15$\mu m$.
 \label{fig7}
\end{figure}

\end{document}